\documentclass[prd,aps,floats,floatfix,superscriptaddress,preprintnumbers,
showpacs,eqsecnum,nofootinbib,twocolumn]{revtex4}
\usepackage{latexsym,array,theorem,mathrsfs,bm,float}
\usepackage{amsfonts,amsmath,amssymb,latexsym,array,afterpage,theorem,mathrsfs,bm,float,epsfig,color,graphicx,tabularx,here,multirow}
\usepackage{subfig}
\usepackage{psfrag}

\newcommand{\bea}{\begin{eqnarray}}
\newcommand{\ena}{\end{eqnarray}}
\newcommand{\be}{\begin{equation}}
\newcommand{\ee}{\end{equation}}

\newcommand{\ma}[1]{\mbox{$\mathcal{#1}$}}

\newcommand{\calhR}[1]{\raisebox{2ex}{\tiny ({\em h})}\hspace{-0.8em}{\ma R}}

\begin{document}

\title{
Fitting rotation curve of galaxies by de Rham-Gabadadze-Tolley massive gravity }


\author{Sirachak {\sc Panpanich}}
\email{sirachakp@gmail.com}
\address{High Energy Physics Theory Group, Department of Physics, 
Faculty of Science, Chulalongkorn University, Phyathai Rd., 
Bangkok 10330, Thailand}
\author{Piyabut {\sc Burikham}}
\email{piyabut@gmail.com}
\address{High Energy Physics Theory Group, Department of Physics, 
Faculty of Science, Chulalongkorn University, Phyathai Rd., 
Bangkok 10330, Thailand}


\date{\today}

\begin{abstract}
We investigate effects of massive graviton on the rotation curves of the Milky Way, spiral galaxies and Low Surface Brightness~(LSB) galaxies. Using a simple de Rham, Gabadadze, and Tolley (dRGT) massive gravity model, we find static spherically symmetric metric and a modified Tolman-Oppenheimer-Volkoff (TOV) equation.  The dRGT nonlinear graviton interactions generate density and pressures which behave like a dark energy that can mimic the gravitational effects of a dark matter halo. We found that rotation curves of most galaxies can be fitted well by a single constant-gravity parameter $\gamma \sim m_{g}^{2}C \sim 10^{-28}~{\rm m^{-1}}$ corresponding to the graviton mass in the range $m_g \sim 10^{-21}-10^{-30} {\rm eV}$ depending on the choice of the fiducial metric parameter $C\sim 1-10^{18}~\text{m}$.  Fitting rotation curve of the Milky Way puts strong constraint on the Yukawa-type coupling of the massive graviton exchange as a result of the shell effects. 
\end{abstract}

\maketitle

\section{Introduction}
\label{intro}

Since the discovery of asymptotically flat rotation curves of most observable galaxies in contradiction to the small amount of visible masses that are much less than gravitationally required, there have been a number of hypotheses proposed to explain this phenomenon.  The lists include the proposal of dark matter halo \cite{Navarro:1995iw,Begeman:1991iy} and modified gravity theories such as the Modified Newtonian dynamics (MOND) \cite{Milgrom:1983ca,Famaey:2011kh}, the $f(R)$ gravity \cite{Capozziello:2006ph} and the two-metric model~\cite{Sporea:2017eph}.  However at the galactic scales, gravitational lensing studies~\cite{Clowe:2003tk,Markevitch:2003at} strongly favour the existence of local distribution of dark matter throughout the space ranging from the galactic to the supercluster scale especially when dynamics of galaxies are involved.

Extending beyond the galactic scale further out, expansion effects of the spacetime start to take over.  At this extragalactic scale, General Relativity~(GR) requires existence of dark energy to explain the accelerated expansion of the Universe.  The two biggest problems in gravitational physics are the asymptotically flat rotation curves of galaxies in the scale of kiloparsecs and the accelerated expansion of spacetime in the extragalactic scale of megaparsecs.  

There are certain classes of modified gravity theories that extend the General Relativity to address cosmological phenomena, a notable one is the massive gravity theory.  After discovery of the gravitational waves from merging black holes/massive stars by LIGO/VIRGO \cite{Abbott:2016blz,TheLIGOScientific:2017qsa} the mass of graviton has been severely constrained, at least in certain straightforward interpretation using dispersion relation. If the inverse mass or Compton wavelength of graviton is of the order of parsec scale (see also \cite{Aoki:2016zgp} for massive graviton being dark matter), then it would be interesting to see effects of the massive graviton on the rotation curves of various types of galaxies; large and small, bright and dim. 

One of the promising massive gravity theories is the nonlinear ghost-free dRGT massive gravity~\cite{deRham:2010ik,deRham:2010kj}. Static spherically symmetric solution in the simplest dRGT model has two additional characteristic scales comparing to the Schwarzschild solution in GR~\cite{Ghosh:2015cva}, $\gamma$ and $\Lambda$~(see Sec.~\ref{effecthalo} for definitions).  These two parameters can be set to address the two problems of dark matter and dark energy at the galactic and extragalactic scales in a single framework of the dRGT model.  Although the dRGT massive gravity has a problem on cosmological solution because of ghost instabilities \cite{DeFelice:2012mx}~(see however \cite{deRham:2014naa,deRham:2014fha,Gumrukcuoglu:2014xba,Heisenberg:2016spl}), its phenomenology is still interesting.    

In this work we explore effects of the massive graviton self-interactions in the dRGT model on rotation curves of the Milky Way, a number of representative spiral galaxies and LSB galaxies by fitting with the observational data without adding any additional dark matter halo profile.  First, the theoretical framework and setup of the dRGT theory are described in Section~\ref{basiceq}. We found that in a sense, the massive graviton ``anisotropic fluid'' behaves like a kind of dark energy~(i.e. its equation of state is $P_{r}=-\rho$, however $P_{r}\neq P_{\theta,\phi}$) which interestingly can mimic the dark matter halo on the galactic scale.  In Sec.~\ref{effecthalo}, since the massive graviton acts as anisotropic fluid and forms a halo, we find a modified TOV equation which leads to a dRGT-generalized circular velocity inside the halo.  The bulge parameters of the Milky Way are then refitted using the de Vaucouleurs profile~\cite{Sofue:2008wt,deVaucouleurs:1948lna} in the presence of the massive graviton halo.  The calculation and fitting results of the Milky Way and spiral galaxies are discussed. Sec.~\ref{effectLSB} explores similar dRGT effects in the representative LSB galaxies.  In Sec.~\ref{thinshell}, we consider consequences of the Yukawa-type coupling induced by the massive graviton exchange.  Since the Yukawa potential generates a non-inverse-square-law force, the force from the outer shell of mass does not exactly cancel out and the force from the inner shell has a repulsive correction term.  The combined effect put strong constraints on the Yukawa coupling on the galactic scale.  Lastly, Sec.~\ref{conclusions} concludes our work.


\section{General Setup}
\label{basiceq}
We start with the dRGT massive gravity action
\begin{equation}
S = \int d^4 x \sqrt{-g} \frac{M_{\rm Pl}^2}{2} \left[R + m_g^2 \mathcal{U}(g,f) \right] + S_m \,,
\end{equation}
where $M_{\rm Pl}$ is the reduced Planck mass, $R$ is the Ricci scalar, $m_g$ is the graviton mass, and $\mathcal{U}$ is self-interacting potential of the gravitons. To avoid the Boulware-Deser ghost the self interactions $U(g,f)$ must be in the following form
\begin{align*}
&\mathcal{U} \equiv \mathcal{U}_2 + \alpha_3 \mathcal{U}_3 + \alpha_4 \mathcal{U}_4 \,, \\
&\mathcal{U}_2 \equiv  [\mathcal{K}]^2 - [\mathcal{K}^2] \,, \\
&\mathcal{U}_3 \equiv [\mathcal{K}]^3 - 3 [\mathcal{K}][\mathcal{K}^2] + 2 [\mathcal{K}^3] \,, \\
&\mathcal{U}_4 \equiv [\mathcal{K}]^4 - 6 [\mathcal{K}]^2 [\mathcal{K}^2] + 3[\mathcal{K}^2]^2 + 8 [\mathcal{K}][\mathcal{K}^3] - 6 [\mathcal{K}^4] \,,
\end{align*}
where the tensor $\mathcal{K}^{\mu}_{\nu}$ is defined as
\begin{equation}
\mathcal{K}^{\mu}_{\nu} \equiv \delta^{\mu}_{\nu} - \sqrt{g^{\mu\lambda} \partial_{\lambda} \varphi^a \partial_{\nu} \varphi^b f_{ab}} \,,
\end{equation}
and $[\mathcal{K}] = \mathcal{K}^{\mu}_{\mu}$ and $(\mathcal{K}^i)^{\mu}_{\nu} = \mathcal{K}^{\mu}_{\rho_1} \mathcal{K}^{\rho_1}_{\rho_2} ... \mathcal{K}^{\rho_i}_{\nu}$. The physical metric is $g_{\mu\nu}$ whereas $f_{\mu\nu}$ is a reference (fiducial) metric and $\varphi^a$ are the St$\Ddot{\rm u}$ckelberg fields.. In this work we use the unitary gauge, $\varphi^a = x^{\mu} \delta^a_{\mu}$, thus
\begin{align*}
\sqrt{g^{\mu\lambda} \partial_{\lambda} \varphi^a \partial_{\nu} \varphi^b f_{ab}} &= \sqrt{g^{\mu\lambda} f_{\lambda\nu}} \,.
\end{align*}
 
Variation with respect to $g^{\mu\nu}$ gives the field equations
\begin{equation}
G^{\mu}_{\nu} + m^2_g X^{\mu}_{\nu} = 8 \pi G T^{\mu (m)}_{\nu} \,.
\end{equation}
$T^{\mu (m)}_{\nu}$ is the energy-momentum tensor obtained from the matter Lagrangian. The massive graviton tensor $X^{\mu}_{\nu}$ is~\cite{Berezhiani:2011mt,Cai:2012db,Ghosh:2015cva} 
given by
\begin{align}
X^{\mu}_{\nu} &= \mathcal{K}^{\mu}_{\nu} - [\mathcal{K}] \delta^{\mu}_{\nu} - \alpha \left[(\mathcal{K}^2)^{\mu}_{\nu} - [\mathcal{K}]\mathcal{K}^{\mu}_{\nu} +\frac{1}{2}\delta^{\mu}_{\nu} ([\mathcal{K}]^2 - [\mathcal{K}^2])\right] \nonumber \\
&~~ + 3 \beta \left[(\mathcal{K}^3)^{\mu}_{\nu} - [\mathcal{K}](\mathcal{K}^2)^{\mu}_{\nu} +\frac{1}{2}\mathcal{K}^{\mu}_{\nu} ([\mathcal{K}]^2 - [\mathcal{K}^2])\right. \nonumber \\
&~~ \left. - \frac{1}{6} \delta^{\mu}_{\nu} ([\mathcal{K}]^3 - 3 [\mathcal{K}][\mathcal{K}^2] + 2[\mathcal{K}^3]) \right] \,,
\end{align}
where $\alpha_3 = \frac{\alpha - 1}{3}$ and $\alpha_4 = \frac{\beta}{4} + \frac{1 - \alpha}{12}$.  The terms of order $\mathcal{O}(\mathcal{K}^4)$ disappear under the fiducial metric ansatz:
\begin{equation}
f_{\mu\nu} = 
  \begin{pmatrix}
    0 & 0 & 0 & 0 \\
    0 & 0 & 0 & 0 \\
    0 & 0 & C^2 & 0 \\
    0 & 0 & 0 & C^2 \sin^2 {\theta}
  \end{pmatrix} \,,
\end{equation} 
where $C$ is a positive constant.

We will find a static and spherically symmetric solution, the generic physical metric can be thus expressed in the form
\begin{equation}
ds^2 = - n(r)dt^2 + \frac{1}{f(r)}dr^2 + r^2 d\theta^2 + r^2 \sin^2\theta d\phi^2 \,.
\end{equation}
Consequently, the field equations become
\begin{widetext}
\begin{align}
-\frac{1}{r^2} + \frac{f}{r^2} + \frac{f^{\prime}}{r} &= m_g^2 \left(\frac{3r - 2C}{r} + \frac{\alpha (3r - C)(r - C)}{r^2} + \frac{3 \beta (r - C)^2}{r^2} \right) -  8 \pi G \rho_{m}(r) \,, \label{g00}\\
-\frac{1}{r^2} + \frac{f}{r^2} + \frac{f n^{\prime}}{r n} &= m_g^2 \left(\frac{3r - 2C}{r} + \frac{\alpha (3r - C)(r - C)}{r^2} + \frac{3 \beta (r - C)^2}{r^2} \right) +  8 \pi G P_{m}(r) \,, \label{g11} \\
\frac{f^{\prime}}{2r} - \frac{f n^{\prime 2}}{4 n^2} + \frac{f^{\prime} n^{\prime}}{4 n} + \frac{f n^{\prime}}{2 r n} + \frac{f n^{\prime\prime}}{2 n} &= m_g^2 \left(\frac{3r - C}{r} + \frac{\alpha (3r - 2C)}{r} + \frac{3 \beta (r - C)}{r} \right) +  8 \pi G P_{m}(r) \,. \label{g22}
\end{align}
\end{widetext}
We see that the massive gravitons can be treated as a fluid where density and pressures depend on the radial coordinate $r$ only.  The self interactions of gravitons generate energy and pressures acting as another source of spacetime curvature in addition to the matter.  Since $X^{\mu}_{\nu}$ contains contribution from cosmological constant~($\delta^{\mu}_{\nu}$ in $\mathcal{K}^{\mu}_{\nu}$) and four St$\Ddot{\rm u}$ckelberg scalars which can be decomposed into helicity 1 and 0 modes~\cite{deRham:2014zqa}, the exotic massive graviton fluid owes its properties to these helicity modes while the usual tensor modes contribute to conventional gravity.  From the equations of motion above, the density and pressures of the massive gravitons can be identified as
\bea
\rho_{g}(r)&=& -\frac{m_{g}^{2}}{8\pi G}\left( \frac{3r-2C}{r}+\frac{\alpha(3r-C)(r-C)}{r^{2}} \right.
\nonumber\\
&&\left.+\frac{3\beta (r-C)^{2}}{r^{2}}\right) = - P_{g}^r (r) \,, \label{rhog} \\
P_{g}^{\theta,\phi}(r)&=&\frac{m_{g}^{2}}{8\pi G}\left( \frac{3r-C}{r}+\frac{\alpha(3r-2C)}{r}+\frac{3\beta (r-C)}{r}\right) \,. \nonumber \\ \label{Ptg}
\ena
The pressures are generically anisotropic with $P^{r}_{g}\neq P^{\theta,\phi}_{g}$, so there is a Poincare stress generated by the massive gravitons~\cite{Burikham:2016cwz,Kareeso:2018xum}.
The dRGT massive graviton behaves more like {\it anisotropic dark energy} with $P_{g}^{r}=-\rho_{g}$.  Interestingly, it can mimic the gravitational effects of a dark matter halo in most kinds of galaxies as to be demonstrated in subsequent sections.


\section{EFFECTS OF A MASSIVE GRAVITON HALO IN THE MILKY WAY AND SPIRAL GALAXIES }
\label{effecthalo}

In this section we will find circular velocity of the Milky Way and a number of representative spiral galaxies from the modified TOV equation in the presence of the dRGT gravitons.  The massive gravitons will play the role of a dark matter halo resulting in the asymptotically flat rotation curves.

From Eq. (\ref{g00}), integrating from $0$ to $r$ we find 
\begin{equation}
f(r) = 1 - \frac{2 G m(r)}{r} - \frac{\Lambda r^2}{3} + \gamma r + \zeta \,, \label{fb}
\end{equation}
where $m(r) \equiv 4 \pi \int_0^r \rho_{m}(r) ~r^2 dr$ is the accumulated matter mass within radius $r$ and
\begin{align}
\Lambda &\equiv - 3 m^2_g (1 + \alpha + \beta) \,, \label{lambda} \\
\gamma &\equiv - m^2_g C (1 + 2 \alpha + 3 \beta) \,, \label{gamma} \\
\zeta &\equiv m_g^2 C^2 (\alpha + 3 \beta) \,.
\end{align}
$\Lambda$ corresponds to the cosmological constant. $\gamma$ and $\zeta$ are constants depending on the graviton mass and other parameters. For $m_g = 0$, the solution reduces to the conventional GR solution.

To obtain flat space with $\zeta =0$, we choose $\alpha=-3\beta$ and require that $\beta=1/2+\epsilon~(1\gg \epsilon >0)$ in order to obtain positive $\Lambda,\gamma$ and finely tune $\Lambda$ to the observed value by the smallness of $\epsilon$.  For this particular choice, the density and pressures of massive graviton given by (\ref{rhog}) and (\ref{Ptg}) take the form
\bea
\rho_{g}&=&-\frac{m_{g}^{2}}{8\pi G}\left(3+2\alpha-\frac{2C}{r}(1+\alpha)\right)=-P_{g}^{r}, \label{rhgr} \\
P_{g}^{\theta,\phi}&=&\frac{m_{g}^{2}}{8\pi G}\left(3+2\alpha-\frac{C}{r}(1+\alpha)\right) \,.
\ena
Some notable features of the massive graviton density is that it can be negative for small $r$ region where the radial pressure is positive, violating energy conditions.  For large $r$, the pressures become negative while the density is positive.  In order to understand the peculiar behaviour of contribution from the massive graviton, we rewrite the density in (\ref{rhgr}) as 
\be
\rho_{g}=\frac{\Lambda}{8\pi G}\left(1-\frac{r_{*}}{r}\right), \label{rhgr1} \\
\ee
where $\Lambda = 6\epsilon m_{g}^{2}, r_{*}=C\displaystyle{\left(1+\frac{1}{6\epsilon}\right)}$.  The positive energy condition is violated at $r<r_{*}$ which could be a very large distance for our fine-tuned choice of $\Lambda \sim 10^{-52}~\text{m}^{-2}$.  For e.g. $m_{g}=6.16\times 10^{-21}~\text{eV}$~(this is the best-fit value from the Milky Way rotation curve as we will see later on), $C=1$ m, the value of $\epsilon$ is roughly $2\times 10^{-26}$ giving $r_{*}\sim 9\times 10^{24}$ m, in the order of Gpc.  However, the value of (negative) $\rho_{g}$ is only about $50$ g/m$^{3}$ at $r=1$ m and continue to increase to approach the cosmological constant value $\rho_{\Lambda}=\Lambda/8\pi G$ for $r\gg r_{*}$.

Even though the negative density region does not totally lie within the horizon of the Supermassive Black Hole~(SMBH) at the center~(the Schwarzschild radius is only $1.2\times 10^{10}$ m) and even extends to intergalactic scale, its negative density $\rho_{g}$ must be added to the energy/matter that made up the black hole and normal matter resulting in the total positive mass.  The total effects of $\rho_{g},P_{g}$ up to $r$ are already integrated into the expression (\ref{fb}) in the last three terms.  On the other hand, since the cutoff scale of effective theory in the dRGT model is $\Lambda_{3}=(m_{g}^{2}M_{P})^{1/3}\sim 7\times 10^{-5}$ eV, the corresponding transition radius~(Vainshtein radius, see e.g. Ref.~\cite{Hinterbichler:2011tt}) between linear and nonlinear regimes is thus $r_{V}\sim \left(\frac{G M}{m_g^2}\right)^{1/3}\sim 10^{13}$ m for the Milky Way mass $M=10^{11}M_{\odot}$.  For galactic distances of order kpc $\gg r_{V}$, the nonlinear effects of the dRGT are expected to be mostly suppressed and the modification on GR becomes apparent.

The mass of SMBH at the center of the Milky Way is $4.1 \pm 0.6 \times 10^6 M_{\odot}$ \cite{Ghez:2008ms} and it is included in the bulge profile of the galaxy.  For the Milky Way, $m(r)$ includes both visible mass of the bulge and the disk.  The cosmological constant $\Lambda$ is $1.11247 \times 10^{-52} ~{\rm m^{-2}}$ which is calculated from the dark energy density of the Universe \cite{Ade:2015xua}. 

Substituting Eq. (\ref{fb}) into Eq. (\ref{g11}) we find
\begin{equation}
\frac{d \ln n}{dr} = \frac{2G m(r) + 8 \pi G P_{m}r^{3} -\frac{2 \Lambda r^3}{3} + \gamma r^2}{r (r - 2 G m(r) -\frac{\Lambda r^3}{3} + \gamma r^2 + \zeta r)} \,. \label{mTOV}
\end{equation}
When $m_g = 0$, i.e. $\Lambda = \gamma = \zeta = 0$, we obtain the usual TOV equation. 
Since a galactic scale is generally a non-relativistic scale, we can use the Newtonian approximation. From the geodesic equation, the acceleration in the radial direction is given by
\begin{equation}
\frac{d^2 r}{dt^2} = \frac{1}{2} \partial_r h_{00} \,, \nonumber
\end{equation}
where $h_{00}$ comes from a small perturbation on metric tensor, i.e. $g_{\mu\nu} = \eta_{\mu\nu} + h_{\mu\nu}$. For an orbital object the circular velocity is given by
\begin{align}
v_c^2 (r) = - \frac{1}{2} r \partial_r h_{00} \,. \label{vc}
\end{align}
We can ignore pressures of the visible matter and assume that gravity is weak. Using Eq. (\ref{vc}), (\ref{fb}) and $n(r)\simeq f(r)$~(from Eqn.~(\ref{g00}) and (\ref{g11}) when $\rho_{m}$ is small) we find 
\begin{equation}
v_{c} (r) = \sqrt{\frac{G m(r)}{r} -\frac{\Lambda r^2}{3} + \frac{\gamma r}{2}} \,. \label{TOV}
\end{equation}
This is the total circular velocity including effects from both the visible mass $m(r)$ and the massive graviton.

To calculate contribution of the bulge to the rotation velocity, we use the de Vaucouleurs law which is a surface-brightness profile of most elliptical galaxies, as the surface mass density profile of the bulge: 
\begin{equation}
\Sigma (R) = \Sigma_b e^{-\kappa \left[\left(\frac{R}{R_b}\right)^{1/4} - 1 \right]} \,,
\end{equation}
where $\kappa = 7.6695$ is a dimensionless constant, $R_b$ is a scale radius, and $\Sigma_b$ is the surface density at $R_b$. Since this profile is applied to a circular plane (two dimensions), we have to reproduce the volume mass density, $\rho(r)$, before calculating $M_{\rm bulge} (r)$. We show detail of calculations in Appendix \ref{deVau}. 

For the disk we use a circular velocity from Ref.~\cite{Sofue:2008wt} without refitting parameters. We argue that in order to provide a good fit on the rotation curve, the massive graviton should yield similar effects as a dark matter halo, thus the parameters on those models should not be changed from Ref.~\cite{Sofue:2008wt} much. Moreover, we assume that $v_{\rm disk}^2 (r) = GM_{\rm disk} (r)/r$ for simplicity. In fact the disk component does not have a spherical symmetry, however for the exponential disk the circular velocity of the thin disk is close to a circular velocity from an equivalent spherical distribution at large distances \cite{Binney}. Therefore, we use
\begin{equation}
\frac{G m(r)}{r} = \frac{G M_{\rm bulge} (r)}{r} + v_{\rm disk}^2 (r) \,. 
\end{equation}

In total, we have three free parameters, they are $\gamma$, $\Sigma_0$, and $R_b$ where the last two parameters come from the de Vaucouleurs profile. The best-fit values of the rotation curve for the Milky Way are 
\begin{eqnarray}
\gamma &=& 4.87739 \times 10^{-28} ~{\rm m^{-1}} \,, \nonumber \\
\Sigma_b &=& 4.45009 \times 10^{39} ~{\rm kg/kpc^2} \,, \nonumber \\
R_b &=& 0.553887 ~{\rm kpc} \,. \nonumber
\end{eqnarray}
If the constant $C$ in the fiducial metric is equal to $1 ~{\rm m}$, we find 
\begin{equation}
m_g = 6.16304 \times 10^{-21} ~{\rm eV} \,.
\end{equation}
Result of the massive graviton halo's fit to the rotation velocity of the Milky Way is shown in Fig. \ref{bulge}.

\begin{figure*}[htp]
        {{\includegraphics[width=0.5\textwidth]{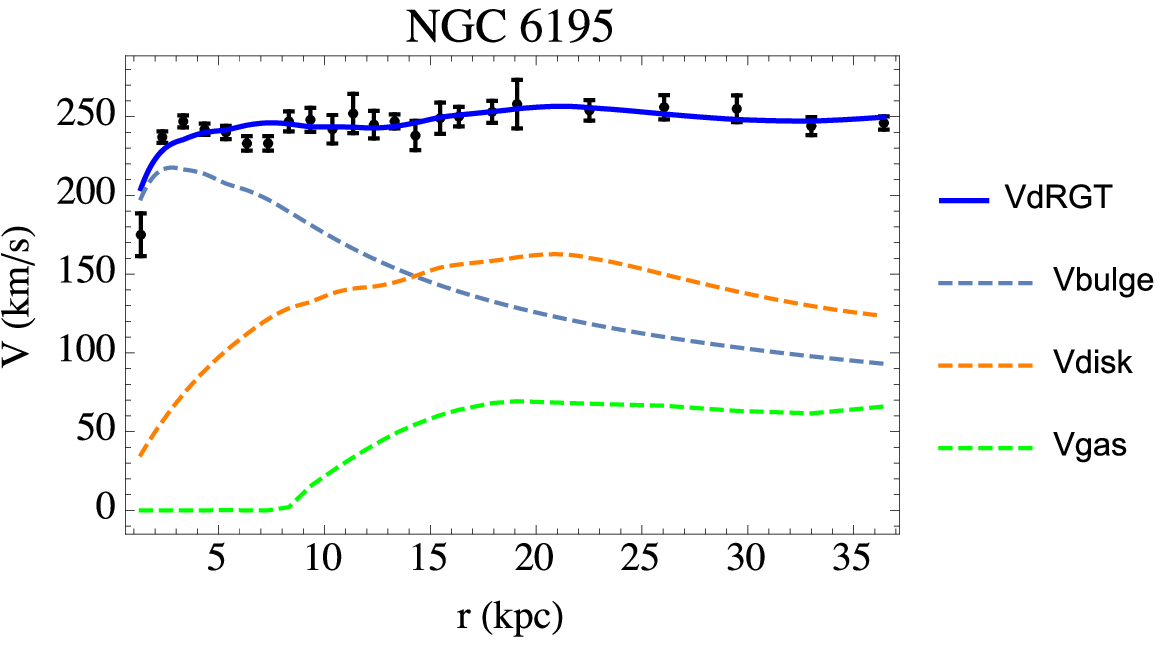}}\hfill
        {\includegraphics[width=0.5\textwidth]{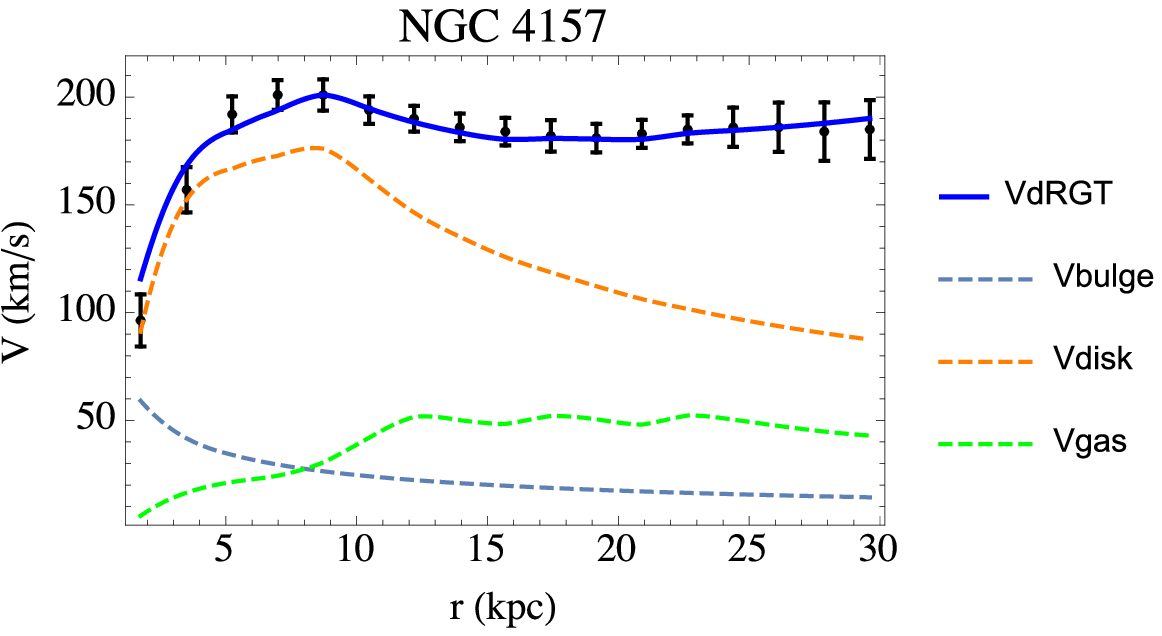}}}\hfill
        {{\includegraphics[width=0.5\textwidth]{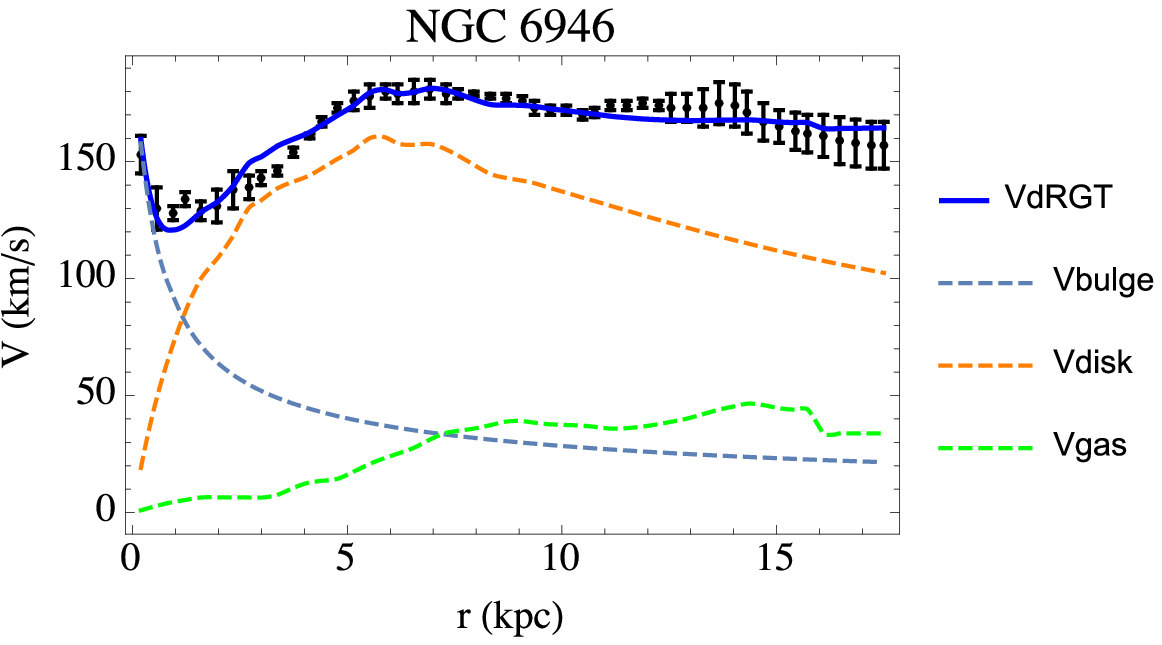}}\hfill
        {\includegraphics[width=0.5\textwidth]{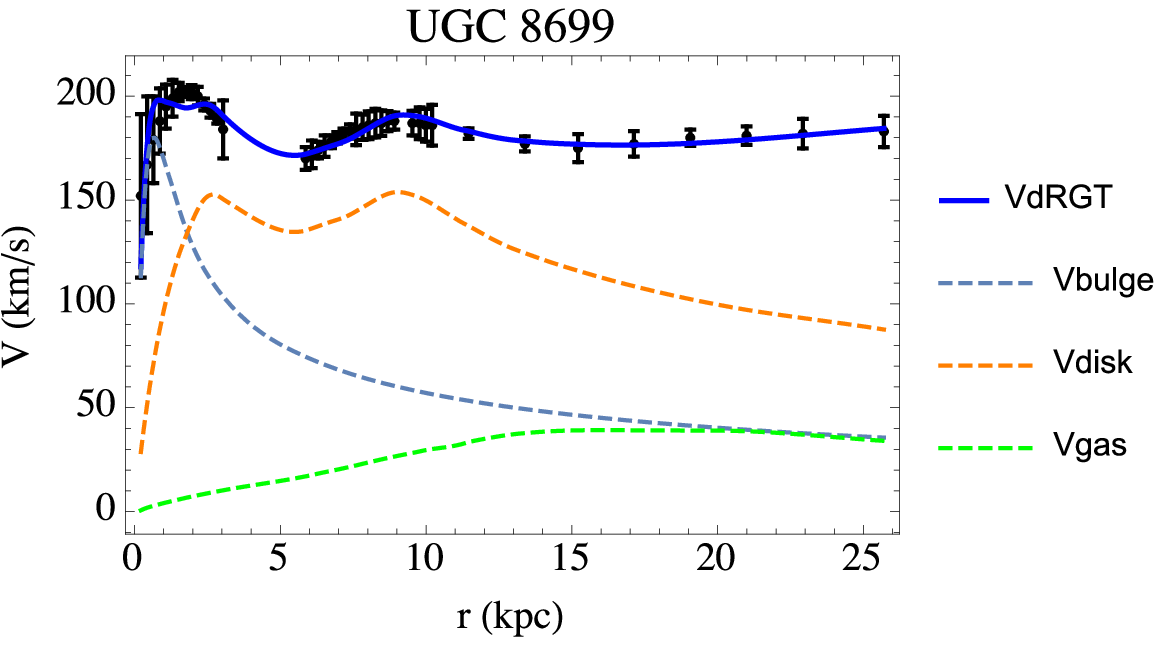}}}\hfill
    \caption{The rotation curve of NGC6195, NGC4157, NGC6946 and UGC8699 with best-fit massive gravity parameters shown in Table \ref{Table S}.  The best-fit curve labelled VdRGT is the circular velocity given by Eqn.~(\ref{TOV}) and (\ref{wspi}). The component Vdisk, Vbulge and Vgas are shown with the weight factor multiplied.}
\label{Spiralfits}
\end{figure*}

In addition to the Milky Way, generically the rotation curves of spiral galaxies can be fitted reasonably well by the dRGT massive gravity model with only single parameter $\gamma$.  The existence of bulge, disk and gas velocities with flexibility in the value of mass-to-light ratios for each contribution~\cite{Swaters} allows the dRGT model to fit with the observed rotation curves of most of the spiral galaxies listed in SPARC~\cite{Lelli:2016zqa}.  For spiral fitting by Eqn.~(\ref{TOV}), we use  
\begin{equation}
\frac{G m(r)}{r} = x v^{2}_{\rm bulge}(r) + y v_{\rm disk}^2 (r) + z v^{2}_{\rm gas}(r), \label{wspi}
\end{equation}
where $x,y$ and $z$ represents the dimensionless mass-to-light ratio of the bulge, disk and gas respectively.  We show the fits of four representative spirals in Figure~\ref{Spiralfits} with the best-fit parameters listed in Table~\ref{Table S}.

\begin{widetext}
\begin{center}
\begin{figure}[htb]
	\includegraphics[width=6in]{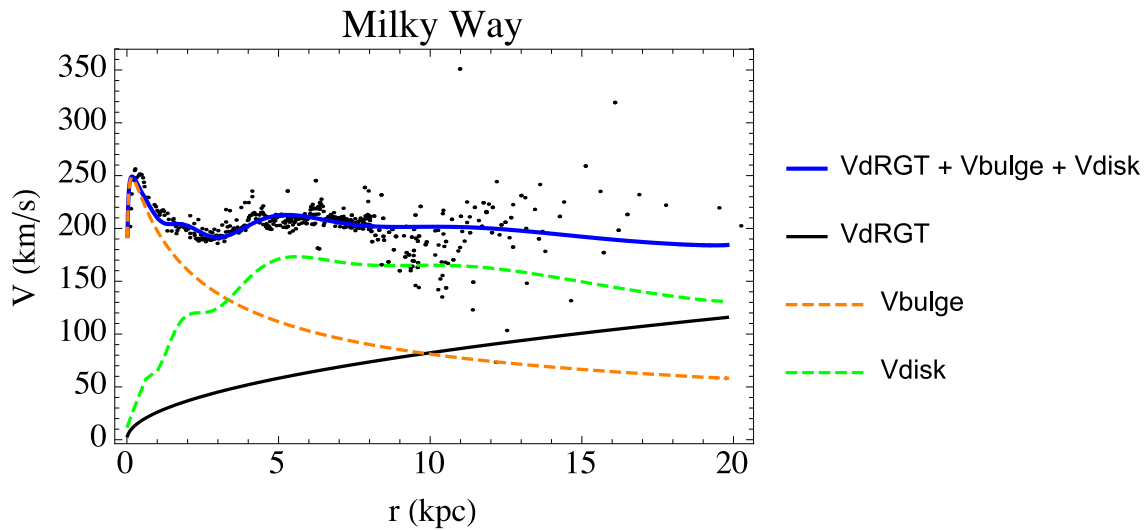} 
	\caption{The rotation curve of the Milky Way. Observational data (black-dot) and circular velocities of the disk (green-dashed) are obtained from Ref.~\cite{Sofue:2008wt}.  The bulge contribution is refitted and presented as the orange-dashed line.  The combination of the bulge, the disk, and the dRGT gravitons is represented in the blue line, whereas the contribution from the dRGT part in the TOV equation (\ref{TOV}) is shown as the black line.} 
\label{bulge}
\end{figure}
\end{center}
\end{widetext}

\begin{widetext}
\begin{table*}[ht]
\begin{center}
  \begin{tabular}{|c||c|c|c|c|}
\hline 
& & & \\[-.5em]
dRGT&$\gamma~(10^{-28}~\rm m^{-1})$&$x,y,z$ &$C~{(\rm m)}$
\\[.5em]
\hline 
\hline
& & & \\[-.6em]
Milky Way&$4.87739$&$1^*,1,0$&$1.00$
\\[1em]
\hline
& & & \\[-.6em]
NGC6195~(Sb)&$6.74171$&$0.7,0.4427,1$&$1.39$
\\[1em]
\hline
& & & \\[-.6em]
NGC4157~(Sb)&$6.43075$&$0.7,0.49216,1$&$1.32$
\\[1em]
\hline
& & & \\[-.6em]
NGC6946~(Scd)&$6.14538$&$0.4580,0.6127,1$&$1.26$
\\[1em]
\hline
& & & \\[-.6em]
UGC8699~(Sab)&$6.70334$&$0.514856,1.18365,1$&$1.38$
\\[1em]
\hline 

 \end{tabular}
    \caption{The fitting parameter $\gamma$ of each spiral galaxy~~(characterized according to Ref.~\cite{deV94}) where $C$ is calculated from $\gamma$ and $m_{g}$ using $\alpha  = -3\beta, \beta = 1/2+\epsilon$ and $m_{g}=6.16304\times 10^{-21}~$ eV.  The weighing factors of $v^{2}_{\rm bulge}, v^{2}_{\rm disk}$, $v^{2}_{\rm gas}$ are $x, y, z$ respectively.  For the Milky Way, the de Vaucouleurs parameters of the bulge are refitted together with the massive gravity parameter $\gamma$ and thus the bulge weight is denoted by $x=1^{*}$. }
\label{Table S}
\end{center}
\end{table*}
\end{widetext}

\section{EFFECTS OF MASSIVE GRAVITON IN THE LSB GALAXIES} 
\label{effectLSB}

\begin{figure*}[htp]
        {{\includegraphics[width=0.5\textwidth]{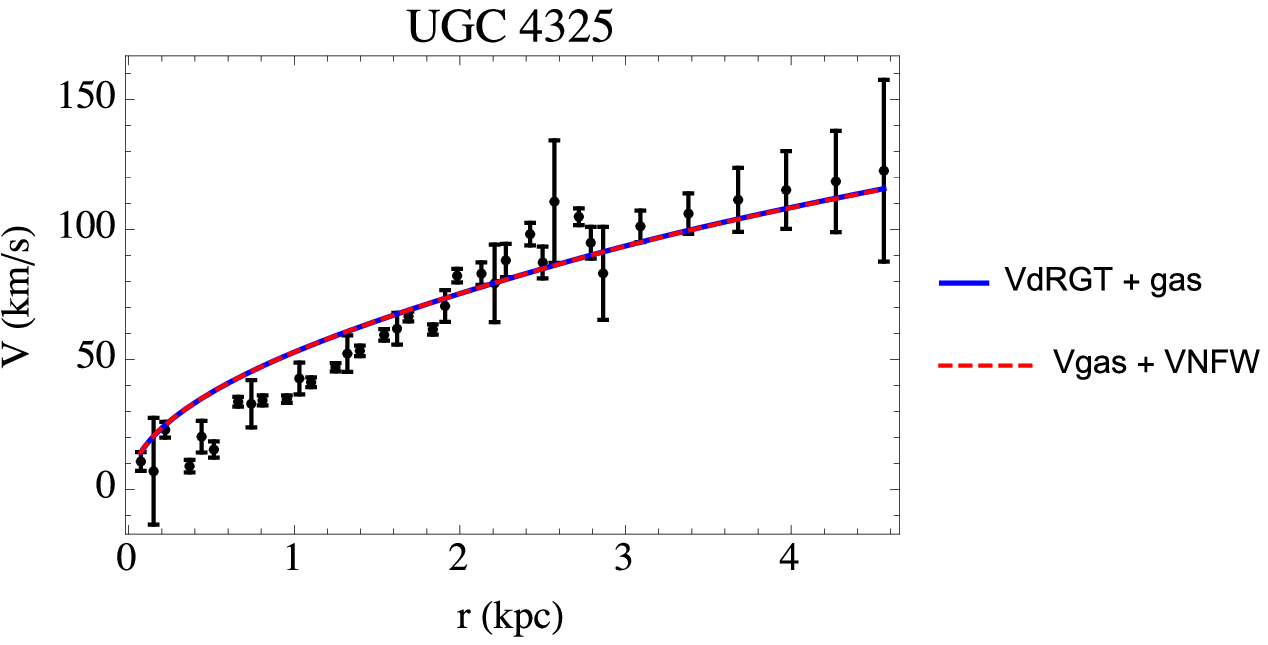}}\hfill
        {\includegraphics[width=0.5\textwidth]{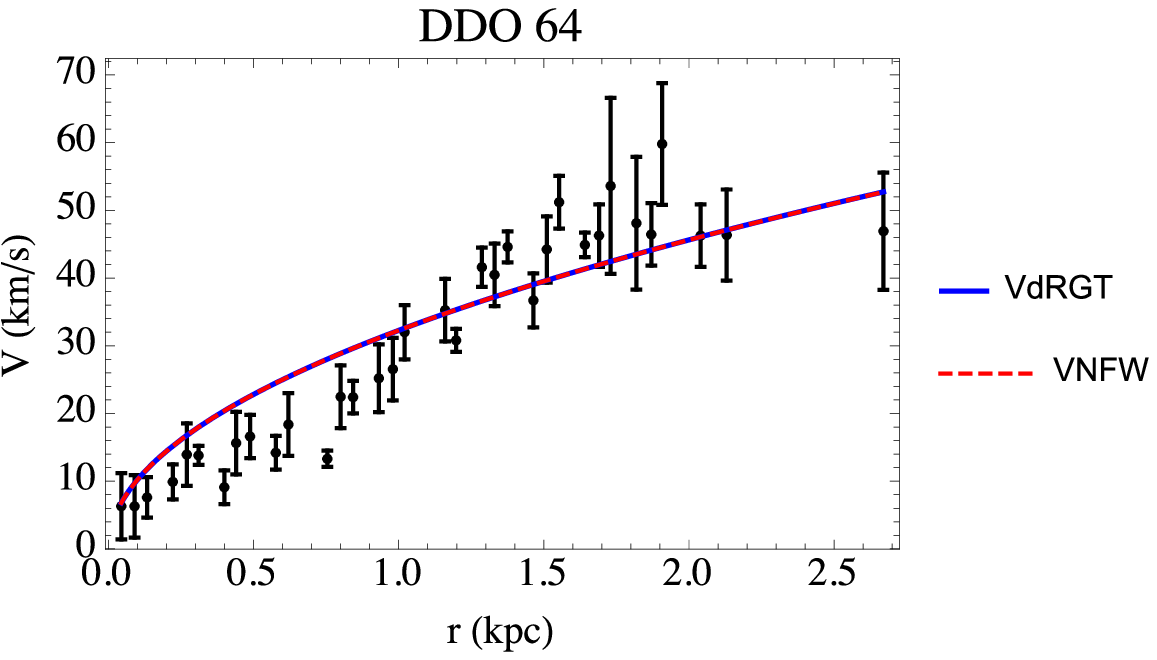}}}\hfill
        {{\includegraphics[width=0.5\textwidth]{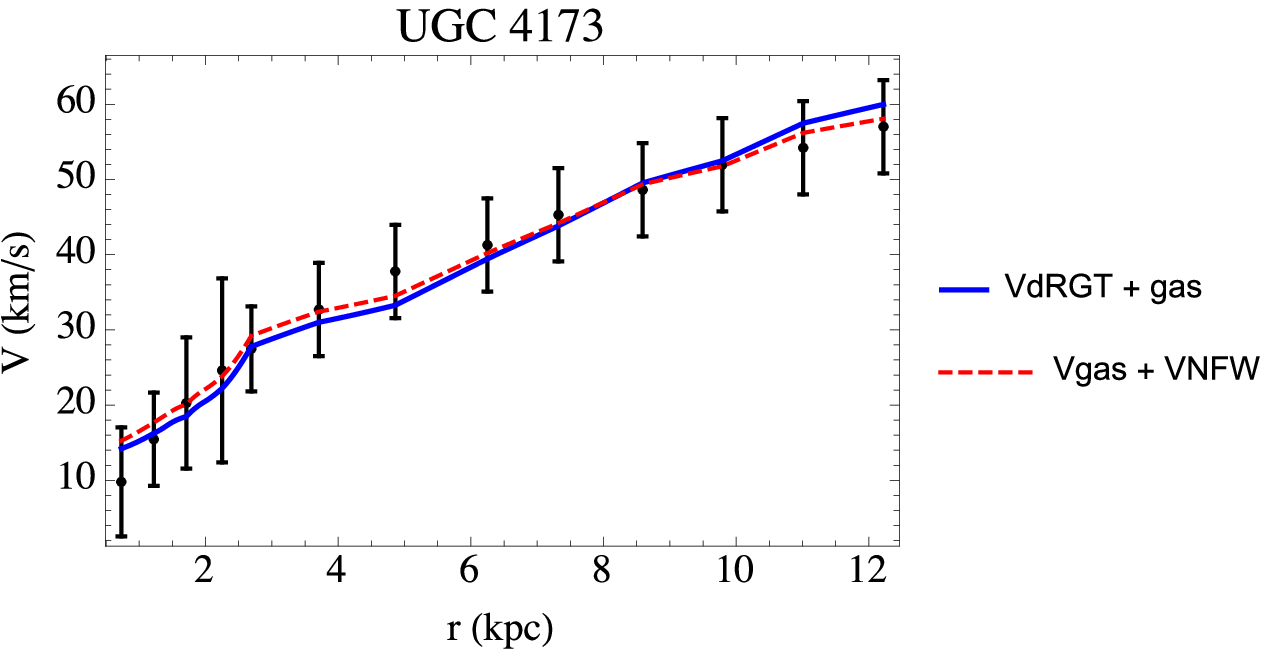}}\hfill
        {\includegraphics[width=0.5\textwidth]{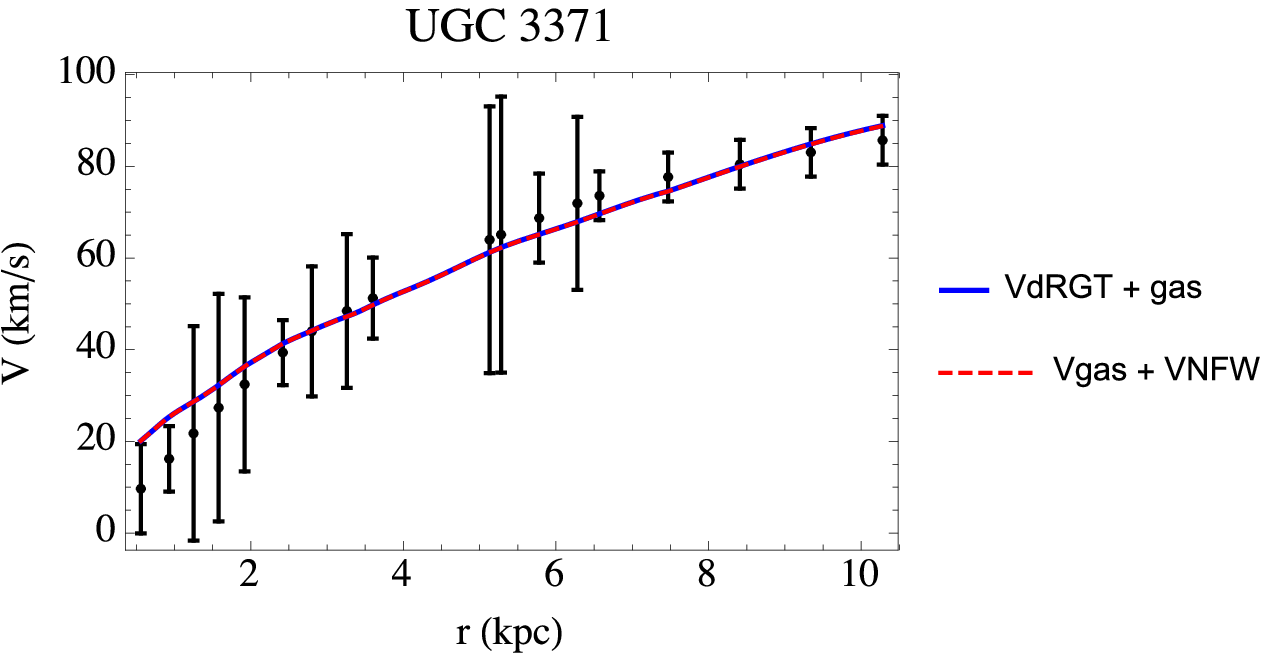}}}\hfill
        {{\includegraphics[width=0.5\textwidth]{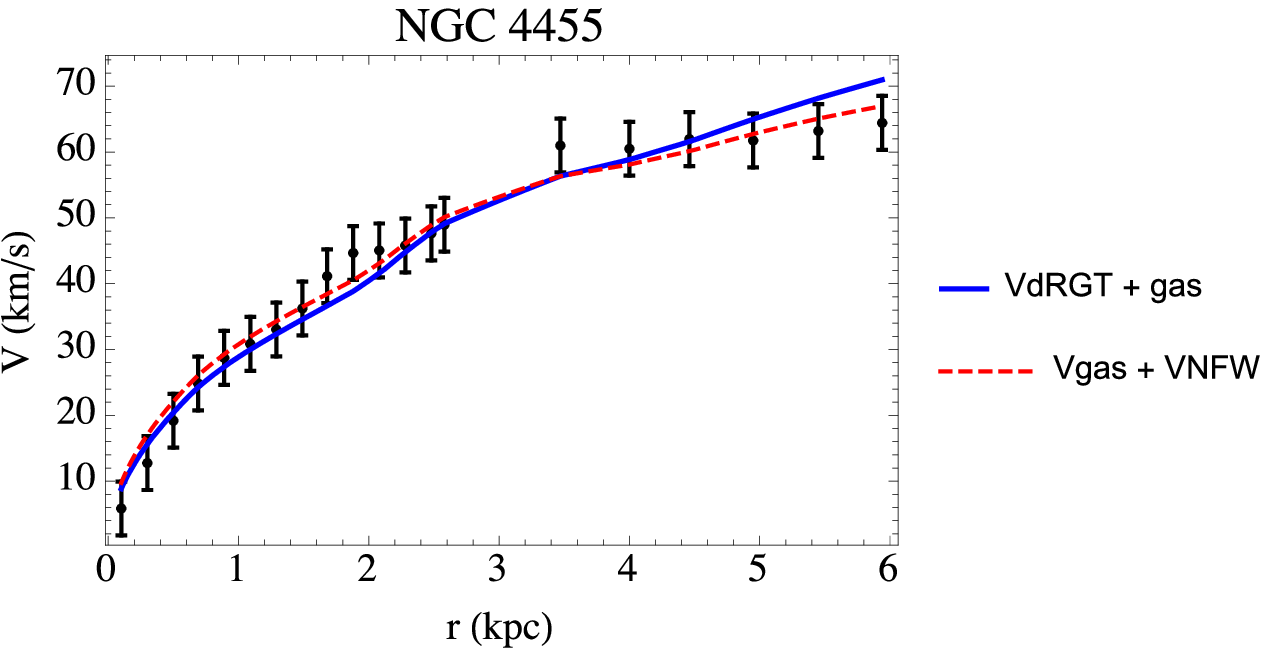}}\hfill
        {\includegraphics[width=0.5\textwidth]{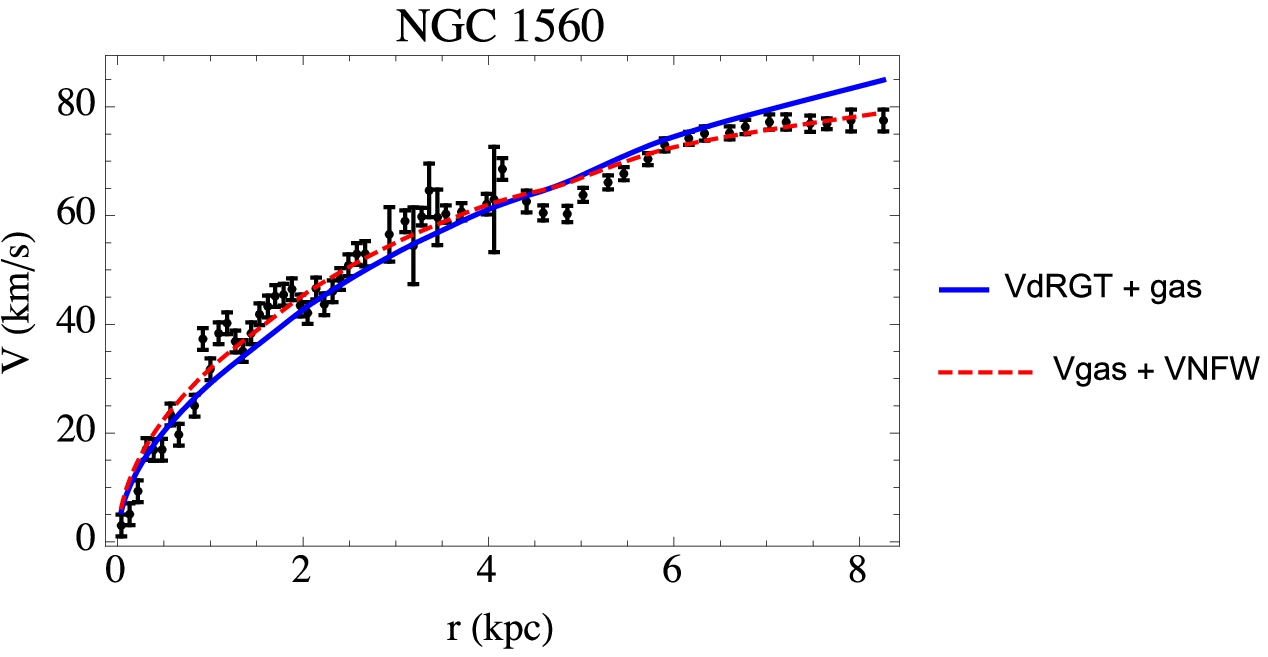}}}\hfill
    \caption{The rotation curve of UGC4325, DDO64, UGC4173, UGC3371, NGC4455 and NGC1560 with best-fit massive gravity parameters shown in Table \ref{Table I}. The blue line represents the fit of dRGT $+$ gas, whereas the red-dashed line represents the fit of NFW dark matter halo $+$ gas. Note that the galaxy DDO64 does not have the gas velocity data, then we fit the plot by pure dark matter halo and dRGT only.}
\label{LSBfits}
\end{figure*}

If the massive graviton ``dark matter'' or massive graviton effects are mainly responsible for the observed rotation speeds of visible matter in the galaxies, we would expect its effect to be seen more in the LSB galaxies where there are less visible matter in proportion.  In this section we explore the possibility of using dRGT massive graviton profile to fit with rotation curves of a number of representative LSB galaxies as shown in Fig.~\ref{LSBfits}.  We assume the contributions from the known matter are given by $v_{\rm gas}$ and $v_{\rm disk}$ from the data files in Ref.~\cite{deBlok:2002vgq,KuziodeNaray:2007qi,deBlokdat}.  The total rotation speed is then
\begin{equation}
v_{\rm tot}(r) = \sqrt{\frac{Gm(r)}{r}-\frac{\Lambda r^2}{3} + \frac{\gamma r}{2}},
\end{equation}
where
\begin{equation}
\frac{Gm(r)}{r} = v_{\rm gas}^{2}(r)+y v_{\rm disk}^{2}(r),
\end{equation}
where $y$ is the mass-to-light ratio of the stellar disk.  Note that for some galaxies, $v_{\rm gas}$ can become negative in the central depression regions due to the effective outward gravity pull, in such case we replace $v_{\rm gas}^{2}(r)\to v_{\rm gas}(r)|v_{\rm gas}(r)|$~\cite{Lelli:2016zqa}.  For most LSB galaxies, we set $y$ to zero, i.e. the minimum disk scenario.  The best-fit parameters are shown in Table~\ref{Table I}~($y=0$ when not shown).  Observe that the values of $\gamma$ are within an order of magnitude of $10^{-28}~\text{m}^{-1}$, the same order as the best-fit value of the Milky Way.  However, a few galaxies, e.g. UGC1230, UGC5005, F5631 and DDO189 require a large $y$ in order to fit decently with the dRGT TOV profile with $\gamma\sim 10^{-28}~{\rm m}^{-1}$.  It should be noted that the best-fit value of $\gamma$ for UGC4325 is roughly one order of magnitude larger than $10^{-28}$ m$^{-1}$.  This is due to the fact that rotation speeds~(at the same distances) of UGC4325 are relatively faster than other LSB galaxies.  For small LSB galaxies, the core-cusp problem~\cite{deBlok:2009sp} remains since the dRGT massive graviton also generates the cusp and not the core (constant)~density in the central region as shown in all plots.  Certainly, we can assume the change in the profile parameters $\alpha, \beta, C$ between the central and far-away region to address the core problem using the dRGT model.  For example, choosing $C=C(r)=k r, \alpha=-3\beta$ in the central region easily gives constant density core $\sim k$~(and negligible cosmological constant part) for the small LSB galaxies.  However, a more complete model of how and what determine the changes in the fiducial metric between regions is definitely required.  In Figure~\ref{LSBfits}, the NFW fits are presented in comparison to the dRGT fits to note the similarity of the two profiles.  The best-fit parameters of the NFW fits are shown in Table~\ref{Table II}.

\begin{figure*}[htp]
        {{\includegraphics[width=0.5\textwidth]{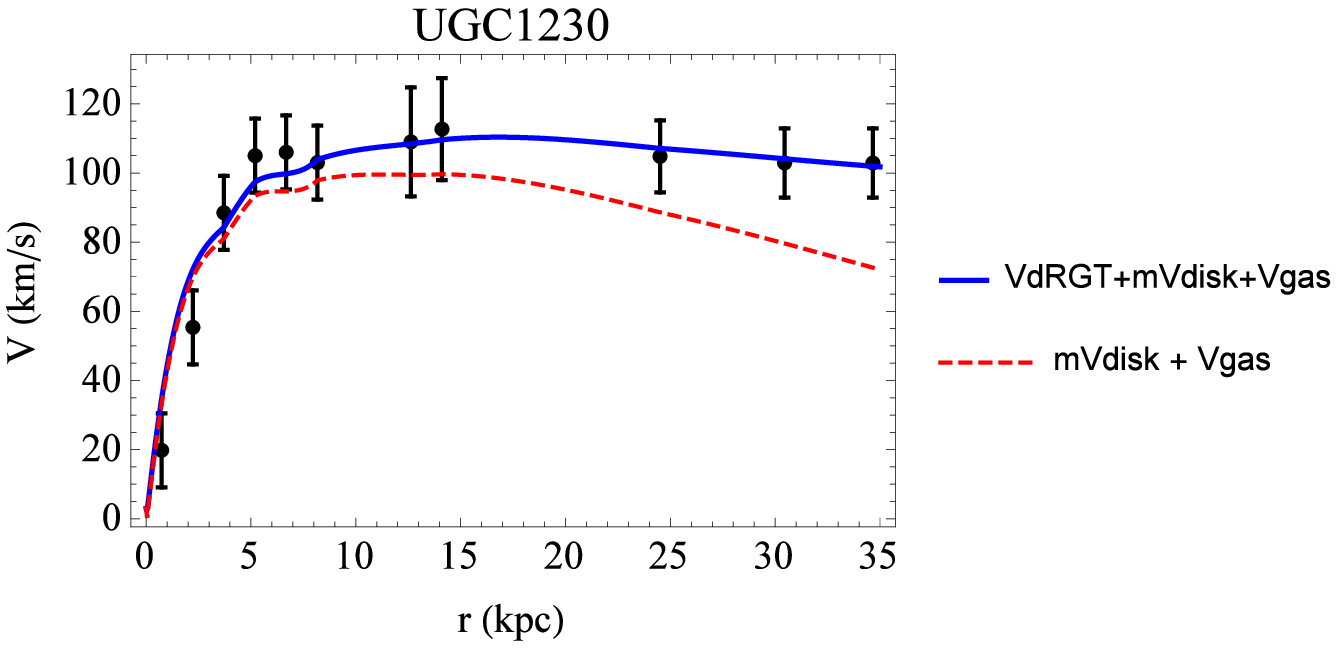}}\hfill
        {\includegraphics[width=0.5\textwidth]{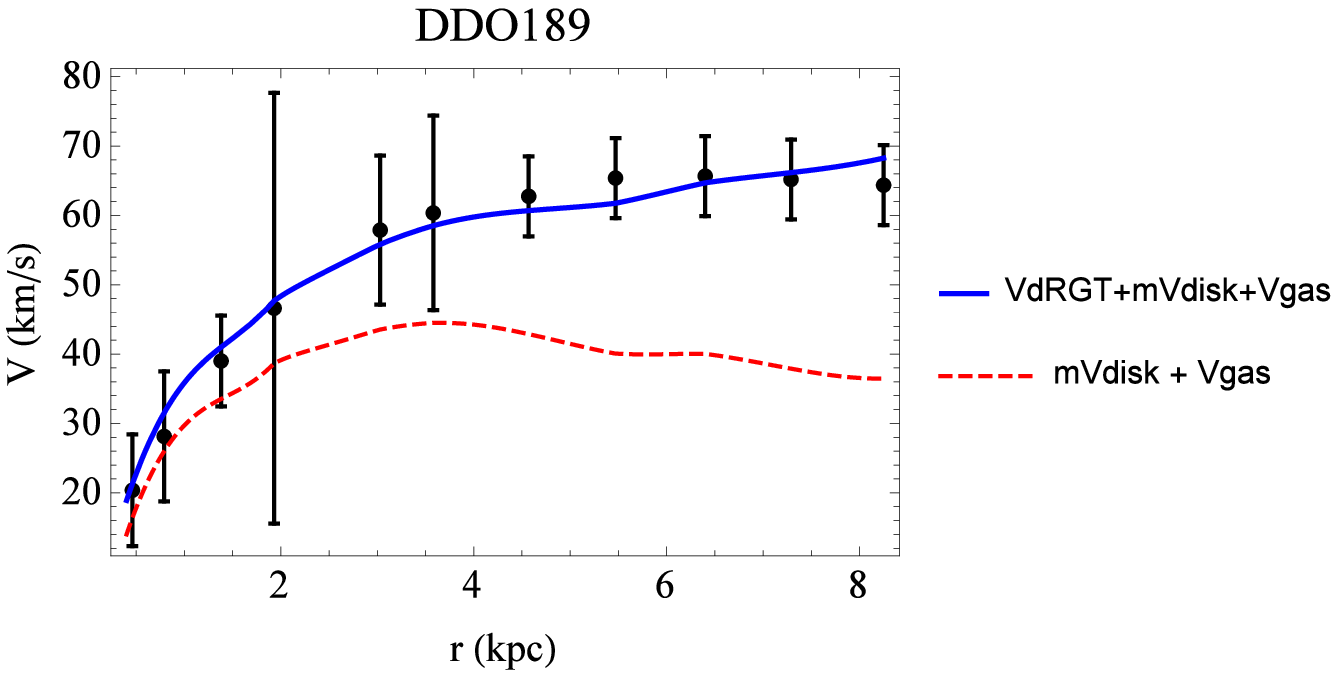}}}\hfill
         {{\includegraphics[width=0.5\textwidth]{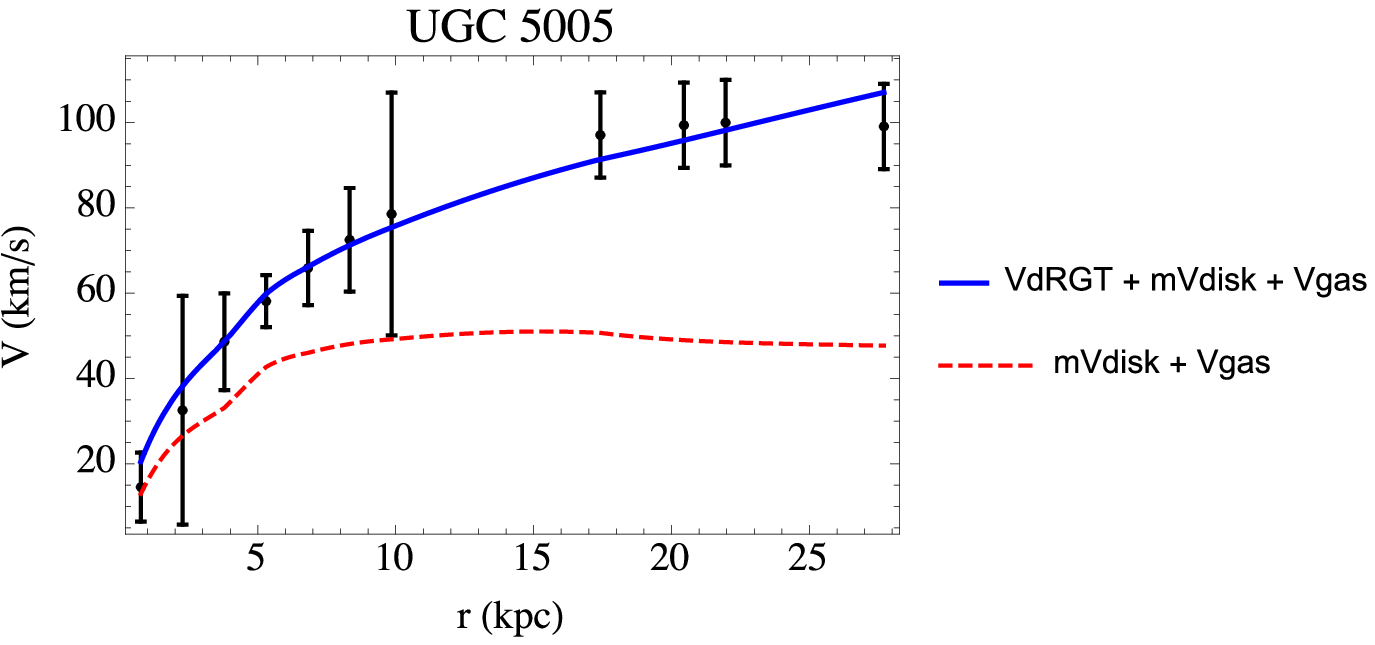}}\hfill
        {\includegraphics[width=0.5\textwidth]{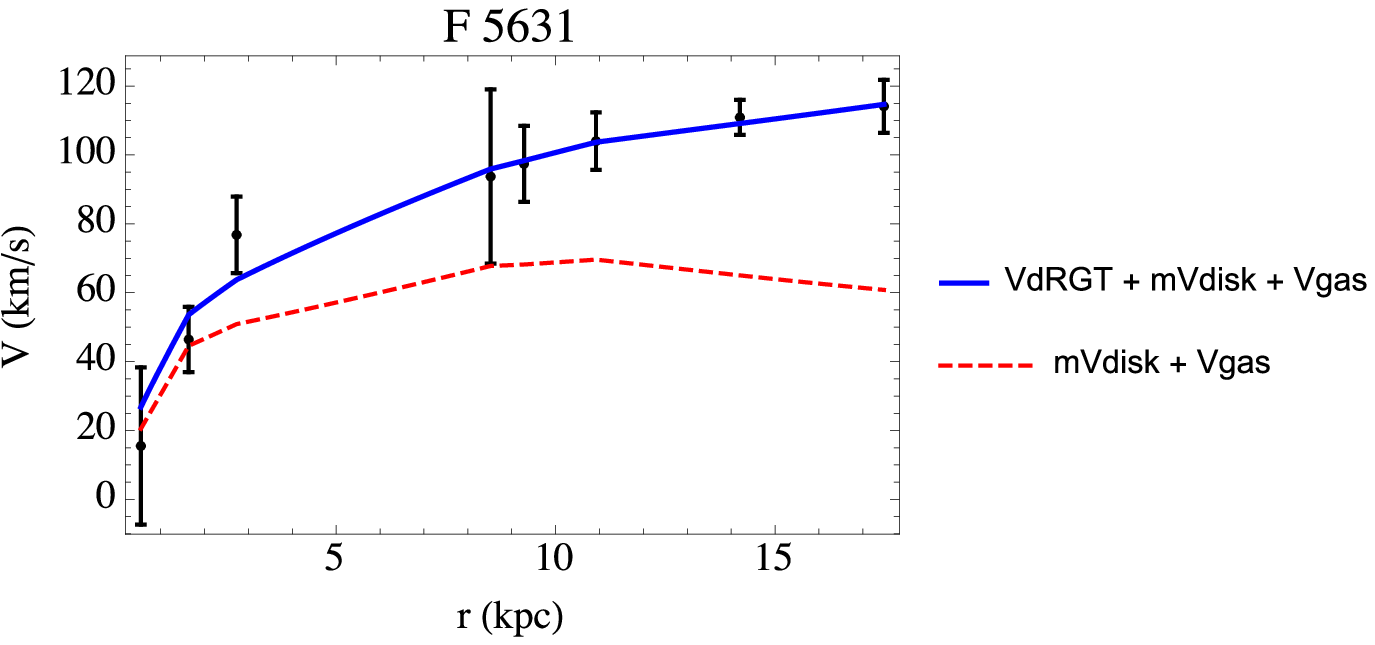}}}\hfill
    \caption{The rotation curve of UGC1230, DDO189, UGC5005 and F5631 with best-fit massive gravity parameters shown in Table \ref{Table I}. The mass-to-light ratios of the stellar disk need to be modified by a factor of $3-10$ for these fits and it is labelled by mVdisk. Massive gravity effects lift the curve up to larger values at large radii. }
\label{LSBfits1}
\end{figure*}

Here we summarize results of LSB galaxies in Fig. \ref{LSBfits}, \ref{LSBfits1} and Table \ref{Table I}.
\begin{table}[H]
\begin{center}
  \begin{tabular}{|c||c|c|c|c|}
\hline
&\multicolumn{2}{|c|}{} &  \multicolumn{2}{|c|}{}\\[-.5em]
dRGT&\multicolumn{2}{|c|}{$\gamma~(10^{-28}~\rm m^{-1})$} &  \multicolumn{2}{|c|}{$C~{(\rm m)}$}
\\[.5em]
\hline 
\hline 
&\multicolumn{2}{|c|}{} &  \multicolumn{2}{|c|}{}\\[-.6em]
UGC4325&\multicolumn{2}{|c|}{$19.9956$} &\multicolumn{2}{|c|}{$4.11$} 
\\[1em]
\hline 
&\multicolumn{2}{|c|}{} &  \multicolumn{2}{|c|}{}\\[-.6em]
DDO64&\multicolumn{2}{|c|}{$7.49968$} &\multicolumn{2}{|c|}{$1.54$} 
\\[1em]
\hline 
&\multicolumn{2}{|c|}{} &  \multicolumn{2}{|c|}{}\\[-.6em]
UGC4173&\multicolumn{2}{|c|}{$1.5165$} &\multicolumn{2}{|c|}{$0.31$} 
\\[1em]
\hline 
&\multicolumn{2}{|c|}{} &  \multicolumn{2}{|c|}{}\\[-.6em]
UGC3371&\multicolumn{2}{|c|}{$5.0024$} &\multicolumn{2}{|c|}{$1.03$} 
\\[1em]
\hline 
&\multicolumn{2}{|c|}{} &  \multicolumn{2}{|c|}{}\\[-.6em]
NGC4455&\multicolumn{2}{|c|}{$5.52132$}&\multicolumn{2}{|c|}{$
1.14$}
\\[1em]
\hline
&\multicolumn{2}{|c|}{} &  \multicolumn{2}{|c|}{}\\[-.6em]
NGC1560&\multicolumn{2}{|c|}{$5.62552$}&\multicolumn{2}{|c|}{$1.16$}
\\[1em]
\hline
&\multicolumn{2}{|c|}{} &  \multicolumn{2}{|c|}{}\\[-.6em]
UGC1230&\multicolumn{2}{|c|}{$1.06442~(y =10.2822)$}&\multicolumn{2}{|c|}{$0.22$}
\\[1em]
\hline
&\multicolumn{2}{|c|}{} &  \multicolumn{2}{|c|}{}\\[-.6em]
DDO189&\multicolumn{2}{|c|}{$2.90571~(y =7.03932)$}&\multicolumn{2}{|c|}{$0.60$}
\\[1em]
\hline
&\multicolumn{2}{|c|}{} &  \multicolumn{2}{|c|}{}\\[-.6em]
UGC5005&\multicolumn{2}{|c|}{$2.39079~(y =2.9603)$}&\multicolumn{2}{|c|}{$0.49$}
\\[1em]
\hline
&\multicolumn{2}{|c|}{} &  \multicolumn{2}{|c|}{}\\[-.6em]
F5631&\multicolumn{2}{|c|}{$3.89396~(y =6.05839)$}&\multicolumn{2}{|c|}{$0.80$}
\\[1em]
\hline
 \end{tabular}
    \caption{The fitting parameter $\gamma$ of each LSB galaxy where $C$ is calculated from $\gamma$ and $m_{g}$ using $\alpha  = -3\beta, \beta = 1/2+\epsilon$ and $m_{g}=6.16304\times 10^{-21}~$ eV.}
\label{Table I}
\end{center}
\end{table}


\section{Thin-Shell Effects of Yukawa force from massive graviton exchange}
\label{thinshell}


We have calculated the Vainshtein radius of the Milky Way in Sect.~\ref{effecthalo}, $r_{V}\sim M^{1/3}\sim 10^{13}$ m.  For dwarf galaxies with $M\sim 10^{6-8}M_{\odot}$, the Vainshtein radius is roughly one order of magnitude smaller, $r_{V}\sim 10^{12}$ m.  Since the Vainshtein radius lies deep within each galaxy, we expect to observe effects of massive graviton fully in the galactic scale.  In addition to the self interactions of massive gravitons acting as the source of gravity considered in previous Sections, another effect of massive graviton is the Yukawa force between matter, e.g. orbiting stars, gases and massive-graviton dark matter, induced from the massive graviton exchange.  It is well known that a mass inside a spherically symmetric shell of mass will feel no net gravity if and only if the force is exactly inverse-square.  Exchange of massive particles generically produces Yukawa-type force per mass whereby in this situation has radial component
\bea
F_{r}&=&-\partial_{r}V(r)=-\partial_{r}\Big( \frac{-a GM e^{-mr}}{r}\Big), \nonumber \\
&=&-\frac{a GM}{r^2}(1+mr)e^{-mr}\simeq -a GM\Big(\frac{1}{r^{2}}-\frac{m^{2}}{2}\Big), \notag\\
\ena
where $a$ represents the interaction strength relative to the conventional massless gravity $G$ and $m$ is the mass of the exchange particle to be identified with $m_{g}$.  We have also approximated $mr\ll 1$ in the last step.  For the inner thin-shell of thickness $\delta r=1/m$, an inverse-square law part of the force will contribute to the total gravity by shifting $G\to (1+a)G$ while the outer shell receives no correction from the $1/r^{2}$ part of the force.  On the other hand, the Yukawa force also produces a constant force $aGMm^2/2$ between the test mass and the inner and outer shells.  For large shells of mass on the galactic scale, the Yukawa force could be very large.

Performing the standard shell integration of forces, the inner and outer thin-shell forces per (test)~mass are
\begin{equation}
\vec{F}_{inner}=\vec{F}_{outer}\simeq \frac{4\pi Ga}{3}\rho r^{2}(\delta r) m^{2}\hat{r}=-\frac{\Delta v^{2}_{c}}{r}\hat{r}, \label{sheqn}
\end{equation}
where $\delta r<1/m$ is the thickness of the shells.  Only the shell's mass in the vicinity of the test mass exert force to the object due to the short-range nature of the Yukawa force.  For our analysis we set $\delta r=1/2m$.  In SI unit, we replace $m\to mc/\hbar$.  Note that the force is {\it outward} with respect to $r$ for positive $a$, i.e. attractive Yukawa potential.  There is no contradiction since the Yukawa potential screens attractive force to be less than the usual inverse-square and thus the correction term becomes repulsive for the inner shell and more attractive to the outward direction for the outer shell.  As a consequence, the rotation speed will {\it slow down} due to the repulsive shell-force effect (See also \cite{Piazza:2003ri}).

As apparent from Eqn.~(\ref{sheqn}), the Yukawa force is sensitive to the density of matter at each position $r$, the higher the density the larger the negative contribution to the rotation speed.  The size of the shell effect is also sensitive to the thickness of the shell determined by the inverse of the graviton mass.  In total, the shell force from Eqn.~(\ref{sheqn}) is proportional to $am$ and consistently vanishing in the zero graviton mass limit.  By assuming the matter density to be approximately the bulge density of the Milky Way given by Eqn.~(\ref{rhobu}), the shell force from the Yukawa-type potential puts constraint on the parameters
\begin{equation}
a\lesssim10^{-9},\quad \frac{a}{\sqrt{C/\text{meter}}}\lesssim 1.3\times 10^{-8}, \label{cont}
\end{equation}
where the best-fit $\gamma=4.89192 \times 10^{-28} ~{\rm m^{-1}}$ at 95\% C.L.  For larger value of $a$, the fit becomes worse and confidence level drops below 95\%~\footnote{Since the shell effects from Yukawa interaction always reduce the rotation speed in the asymptotically far region, it will always make the fit worse.  The constraint we put here is when this effect starts to make the fit at given $\gamma, \Sigma_{b}, R_{b}$ having worse statistics than 95\% C.L. }.

Constraint (\ref{cont}) gives strong limit on the Yukawa coupling, $a\lesssim 10^{-9}$ for $C<0.01$ meter. This constraint is somewhat consistent with the current limits on the Yukawa coupling of gravity~\cite{Murata:2014nra,Adelberger:2009zz}.  It should be noted that there is no direct constraint on $C$ in the galactic scale, we can choose $C$ to be very large, e.g. $10^{18}$ meters with the compensation of smaller $m_{g}$ by a factor of $10^{9}$~(since $\gamma \sim C m_{g}^{2}$) and the second constraint from (\ref{cont}) on $a$ will simply be redundant.  For this choice, the graviton mass $m_{g}$ will be in the order of $10^{-30}$~eV, the reduced Compton wavelength $\lambdabar_{g}=\hbar/m_{g}c\sim 10^{23}~{\rm m}$, saturating the current limit from the Lunar Laser Ranging Experiments~\cite{Merkowitz:2010kka, deRham:2016nuf}.


\section{Conclusions} 
\label{conclusions}


In this work, the dRGT massive gravity model is fitted with the rotation curves of the Milky Way, representative spiral and LSB galaxies without the requirement of additional dark matter.  The best-fit parameter $\gamma$ has values in the order of $10^{-28}\text{m}^{-1}$ for all galaxies.  We note the similarity of the dRGT rotation curves with those from the NFW profile whilst the dRGT has only single free parameter $\gamma$ to fit. Using rotation curve of the Milky Way, we also put severe constraint $a<10^{-9}$ on the massive graviton Yukawa-type coupling which inevitably generates the shell forces.  The mass of the massive graviton could lie within $10^{-21}-10^{-30}$ eV range depending on the choice of the fiducial metric parameter $C\sim 1-10^{18}$ meters.  Letting $C$ dependent on the position~(see e.g. \cite{Katsuragawa:2015lbl}) could possibly explain much wider range of dark matter effects in small and large galaxies, especially the small LSB galaxies with the constant-density core.   

Finally, if we extend out to the region of galaxy cluster, e.g. the Coma cluster with total mass around $10^{45}$ kg and size $\sim 2$ Mpc, the corresponding value of $\gamma$ that can explain the velocities of galaxies around the center of the cluster is remarkably $\gamma \sim \displaystyle{\frac{2GM}{r^{2}}}\sim 10^{-28}$ m$ู^{-1}$.  The same order of magnitude of $\gamma$ fits to dark matter effect even at the cluster scale.

\section*{Acknowledgement}

We would like to thank a referee for valuable and useful comments that lead to better version of the paper.  S.P. is supported by Rachadapisek Sompote Fund for Postdoctoral Fellowship, Chulalongkorn University.  P.B. is supported in part by the Thailand Research Fund~(TRF), Office of Higher Education Commission (OHEC) and Chulalongkorn University under grant RSA6180002.

\appendix

\section{de Vaucouleurs profile} 
\label{deVau}
The de Vaucouleurs profile \cite{deVaucouleurs:1948lna} is a surface-brightness profile which gives a surface mass density when multiplied by the mass-to-light ratio. The surface mass density profile is then assumed to take the following form
\begin{equation}
\Sigma (R) = \Sigma_b e^{-\kappa \left[\left(\frac{R}{R_b}\right)^{1/4} - 1 \right]} \,,
\end{equation}
where $\kappa = 7.6695$, $R_b$ is a scale radius, and $\Sigma_b$ is the surface density at $R_b$. Since this surface density comes from projection of a spherical density $\rho(r)$ (three dimensions) onto a disk $\Sigma(R)$ (two dimensions), then
\begin{equation}
\Sigma(R) = 2 \int^{\infty}_{R} \rho(r) \frac{r}{\sqrt{r^2 - R^2}} dr \,.
\end{equation}
The $R$ is a distance from center of a disk, whereas $r$ is a distance from center of a spherical object. Using the Abel integral \cite{Binney}, we can calculate the volume mass density as
\begin{equation}
\rho(r) = - \frac{1}{\pi} \int_r^{\infty} \frac{d \Sigma(R)}{dR} \frac{dR}{\sqrt{R^2 - r^2}} \,.  \label{rhobu}
\end{equation}
Then,
\begin{equation}
\rho(r) = \frac{e^{\kappa} \left(\frac{1}{R_b}\right)^{1/4} \kappa \Sigma_b G_{0,8}^{8,0}\left(\frac{r^2 \kappa ^8}{16777216 R_b^2}|
\begin{array}{c}
 0,\frac{1}{8},\frac{1}{4},\frac{3}{8},\frac{3}{8},\frac{1}{2},\frac{5}{8},\frac{3}{4} \\
\end{array}
\right)}{32 \pi ^4 r^{3/4}} \,,
\end{equation}
where $G_{0,8}^{8,0}$ is the Meijer G function. The mass distribution of the bulge is then given by
\begin{widetext}
\begin{align}
M_{\rm bulge}(r) &= 4 \pi \int^r_0 \rho(r) r^2 dr = \frac{e^{\kappa } r^{9/4} \left(\frac{1}{R_b}\right)^{1/4} \kappa \Sigma_b G_{1,9}^{8,1} \left(\frac{r^2 \kappa ^8}{16777216 R_b^2}|
\begin{array}{c}
 -\frac{1}{8} \\
 0,\frac{1}{8},\frac{1}{4},\frac{3}{8},\frac{3}{8},\frac{1}{2},\frac{5}{8},\frac{3}{4},-\frac{9}{8} \\
\end{array}
\right)}{16 \pi ^3} \,.
\end{align}
\end{widetext}
We use this mass distribution to find mass of the bulge of the Milky Way in Sec.~\ref{effecthalo} where we have two fitting parameters, $\Sigma_b$ and $R_b$.

\section{NFW DARK MATTER PROFILE}
\label{NFW}

For six LSB galaxies in Fig.~\ref{LSBfits}, we also use the Navarro-Frenk-White (NFW) profile \cite{Navarro:1995iw} to fit the circular velocity in order to demonstrate the similarity between the NFW and the dRGT profile.  For completeness, we summarize the NFW model here. 

Generically the NFW density profile  can be expressed as
\begin{equation}
\rho_{\rm NFW} = \frac{\rho_i}{\displaystyle{\frac{r}{r_s}}\left(1 + \displaystyle{\frac{r}{r_s}}\right)^2} \,.
\end{equation}
$\rho_i$ is related to the density of the Universe, and $r_s$ is a characteristic radius of the halo. Usually these parameters are expressed in another form by the virial theorem, they are $c = r_{200} / r$ and $V_{200}$. $c$ is a concentration parameter, and $r_{200}$ is the virial radius at which the average density of the halo is equal to $200$ times of the average density of the Universe, and the $V_{200}$ is the circular velocity at $r_{200}$. The circular velocity of the NFW halo is then given by
\begin{equation}
v_{\rm NFW} (r) = V_{200} \sqrt{\frac{1}{x}\frac{\ln (1+ c x) - cx/ (1+x)}{\ln (1+c) - c/ (1+c)}} \,,
\end{equation}
where $x = r/r_{200}$. 

Therefore, the circular velocity for LSB galaxies is
\begin{equation}
v_{\rm tot} (r) = \sqrt{v_{\rm gas}^{2}(r) + v_{\rm NFW}^2 (r)} \,,
\end{equation}
where $v_{\rm gas}$ are obtained from Ref. \cite{deBlok:2002vgq,deBlokdat}. The fitting parameters of galaxies are shown in Table \ref{Table II}.

\begin{table}[H]
\begin{center}
  \begin{tabular}{|c||c|c|c|c|}
\hline
&\multicolumn{2}{|c|}{} &  \multicolumn{2}{|c|}{}\\[-.6em]
NFW&\multicolumn{2}{|c|}{$V_{200}~(\rm km/s)$} &  \multicolumn{2}{|c|}{$c$}
\\[1em]
\hline
\hline
&\multicolumn{2}{|c|}{} &  \multicolumn{2}{|c|}{}\\[-.6em]
UGC4325&\multicolumn{2}{|c|}{$806.588$} &\multicolumn{2}{|c|}{$2.20872$} 
\\[1em]
\hline 
&\multicolumn{2}{|c|}{} &  \multicolumn{2}{|c|}{}\\[-.6em]
DDO64&\multicolumn{2}{|c|}{$673.755$} &\multicolumn{2}{|c|}{$0.829035$} 
\\[1em]
\hline 
&\multicolumn{2}{|c|}{} &  \multicolumn{2}{|c|}{}\\[-.6em]
UGC4173&\multicolumn{2}{|c|}{$74.4$} &\multicolumn{2}{|c|}{$2.0$} 
\\[1em]
\hline 
&\multicolumn{2}{|c|}{} &  \multicolumn{2}{|c|}{}\\[-.6em]
UGC3371&\multicolumn{2}{|c|}{$814.7$} &\multicolumn{2}{|c|}{$0.1$} 
\\[1em]
\hline 
&\multicolumn{2}{|c|}{} &  \multicolumn{2}{|c|}{}\\[-.6em]
NGC4455&\multicolumn{2}{|c|}{$86.8$}&\multicolumn{2}{|c|}{$5.4$}
\\[1em]
\hline
&\multicolumn{2}{|c|}{} &  \multicolumn{2}{|c|}{}\\[-.6em]
NGC1560&\multicolumn{2}{|c|}{$92.9$}&\multicolumn{2}{|c|}{$5.4$}
\\[1em]
\hline
 \end{tabular}
    \caption{The fitting parameters of the dark matter halo of each galaxy from \cite{deBlok:2002vgq}. Except galaxies UGC4325 and DDO64 we refit parameters after including data from \cite{KuziodeNaray:2007qi} with the updated Hubble parameter $H_0=67.8 ~{\rm km/s/Mpc}$.}
\label{Table II}
\end{center}
\end{table}

\end{document}